\documentclass[usenatbib]{mn2e}

\usepackage{graphicx}

\title[]{Discovery of pulsational line profile variations in the $\delta$\,Scuti star
HD\,21190 and in the Ap Sr star HD\,218994\thanks{Based on observations collected at the
European Southern Observatory, Paranal, Chile (ESO programm 78.D-0341(A)).}}

\author[J. F. Gonz\'alez et al.]{J. F. Gonz\'alez$^{1}$\thanks{E-mail: fgonzalez@casleo.gov.ar}, 
S. Hubrig $^{2}$, D.W. Kurtz $^{3}$, V. Elkin $^{3}$, and I. Savanov $^{4}$ \\
$^{1}$Complejo Astron\'omico El Leoncito, Casilla 467, 5400 San Juan, Argentina\\
$^{2}$European Southern Observatory, Casilla 19001, Santiago 19, Chile \\
$^{3}$Centre for Astrophysics, University of Central Lancashire, Preston PR1\,2HE, UK\\
$^{4}$Armagh Observatory, College Hill, Armagh BT61 9DG, Northern Ireland}

\begin{document}



\maketitle

\label{firstpage}

\begin{abstract}
Asteroseismology has the potential to provide new insights into the physics of stellar
interiors. 
We have obtained UVES high time resolution observations of the $\delta$ Scuti star
HD\,21190 and of
the Ap Sr star HD\,218994 to search for pulsational line profile variations.
We report the discovery of a new roAp star, HD\,218994, with a pulsation period of 14.2\,min. This
is one of the most evolved roAp stars. No rapid pulsations have been found in the spectra
of the cool Ap star -- $\delta$ Scuti star HD\,21190. However, we detect with unprecedented
 clarity for a $\delta$\,Sct star moving peaks
in the cores of spectral lines that
indicate the presence of high degree non-radial pulsations in this star. 
\end{abstract}


\section{Introduction}

Asteroseismology has the potential to provide new insights into the physics of stellar interiors.
Among the most promising objects that can be studied through this technique are the rapidly
oscillating Ap (roAp) stars. These pulsate in high-overtone, low-degree, nonradial $p$-modes,
with periods in the range $5.65 - 21.2$\,min. In a previous study Hubrig et al. (2000) discussed
the relationship between the roAp stars and the non-oscillating Ap (noAp) stars and concluded
that the noAp stars are, in general, slightly more evolved than the roAp stars. Cunha (2002)
predicted the theoretical edges of the instability strip appropriate to roAp stars and compared
them with the observations. According to her predictions, stars that are more luminous and/or more
evolved should pulsate with longer periods in the $20 - 25$-min range, and, as a result,
their oscillations might have escaped detection in photometric surveys that have had low sensitivity
to periods of this order. Freyhammer et al. (in preparation) have studied 9 evolved Ap stars near
the terminal age main sequence with high spectral resolution, high time resolution spectra obtained
with the Ultraviolet and Visual Echelle Spectrograph (UVES) on the Very Large telescope (VLT).
At precisions less than 60\,m\,s$^{-1}$ they found no pulsation in any of the nine stars,
lending support to Hubrig et al.'s (2000) conclusion that that noAp stars are somewhat more evolved
than roAp stars. Nevertheless, there is one known evolved roAp star with a 21.2-min period,
HD\,116114 (Elkin et al. 2005), so we do know that more evolved Ap stars can pulsate.

It is important to our understanding of the driving mechanism for roAp star pulsations to find
what physical parameters govern which stars pulsate, and which do not. A current working hypothesis
is that the pulsations are driven mostly by H ionisation
(see, e.g., Balmforth et al. 2001; Saio 2005).
The models of Saio (2005) lead to a clear prediction that lower radial overtone pulsation modes
typical of $\delta$\,Sct stars are suppressed by the magnetic field in Ap stars, hence should
not be found amongst the roAp stars. This is in general agreement with observations:
at present there is no known magnetic $\delta$\,Sct star. There are $\delta$\,Sct stars that are
classified as Ap stars, but at classification dispersions
magnetic Ap stars and non-magnetic Am stars can be confused, so that measurements of the magnetic
fields are   independently needed for $\delta$\,Sct stars classified as Ap stars.

The most evolved star known that is classified as an Ap star is HD\,21190 (Koen et al. 2001) which
is also a known $\delta$\,Sct star. In this paper we discuss our search for pulsational line profile
variations in high time resolution UVES spectra of HD\,21190 to test if it is an roAp star,
as well as a $\delta$\,Sct star. We also give a more detailed analysis of the discovery of
roAp pulsations in the Ap\,Sr star HD\,218994, which was included in the sample of non-pulsating
binary Ap stars in the study Hubrig et al. (2000). Both of these stars are unusual,
hence important: HD\,218994 is one of the most luminous roAp stars, hence the discovery of pulsation
makes it a potential test of Cunha's (2002) theoretical instability strip. HD\,21190 is
the only cool Ap star that is also a $\delta$\,Sct
star. Whether it is magnetic is still to be determined. Our observations show it to be
the best known example of a $\delta$\,Sct star with moving bumps in its line profiles
characteristic of high degree pulsation -- making it a prime target for further study.

\section{Observations}

The spectroscopic time series for HD\,21190 and HD\,218994 have been obtained on November 8 and November 3 2006,
respectively, at ESO with the VLT UV-Visual Echelle Spectrograph UVES at UT2.
We used the UVES DIC2 standard setting covering the spectral
range from 3300\,\AA{} to 4500 \,\AA{} in the blue arm and from 5700\,\AA{} to 7600\,\AA{} in the red arm.
The slit width was set to $0\farcs{}3$ for the red arm, corresponding to a resolving power of
  $\lambda{}/\Delta{}\lambda{} \approx 1.1\times10^5$. For the blue arm, we used
  the slit width of $0\farcs{}4$ to achieve a resolving power of
  $\approx 0.8\times10^5$.
Further, we used the fast readout mode (625kHz, 1x1, low) to keep readout and overhead time to about 30 s.
For both stars we used  exposure times of 3 min, obtaining  time series with a resolution of about 3.7 min.
The spectra were reduced by the UVES pipeline Data Reduction Software (version 2.5; Ballester
et al. 2000) and using standard IRAF routines.
The signal-to-noise ratio of the obtained UVES spectra range
from 40 in the near UV (3500\,\AA{}) to 150--200 in the visual region (4000--7000\,\AA{}).

\section{The $\delta$\,Sct star HD\,21190}

HD\,21190 is a known $\delta$\,Sct star with a variability period of 3.6\,h, discovered by the
Hipparcos mission. Koen et al. (2001) determined the spectral type to be F2\,III SrEuSi, making
it the most evolved Ap star known. According to Cunha (2002), longer period oscillations (20-25 min)
in magnetic roAp stars should exist in the more evolved stars. To search for pulsational line
profile variations on time scales shorter than typical $\delta$\,Sct star pulsations
(of the order of hours), we obtained for this star 14 UVES spectra covering about 50 min,
 corresponding to $\sim$$\frac{1}{4}$  of the known $\delta$\,Sct pulsation period.
Using Hipparcos and ASAS (Pojmanski 2002) photometric databases, we re-calculated the
ephemeris and found that our observations correspond to the phase interval of decreasing brightness.

\begin{figure*}
\label{fig.hd21190}
  \begin{center}
    \begin{minipage}[t]{8.5cm}
    \includegraphics*[width=9cm,height=7cm, bb=130 240 540 552]{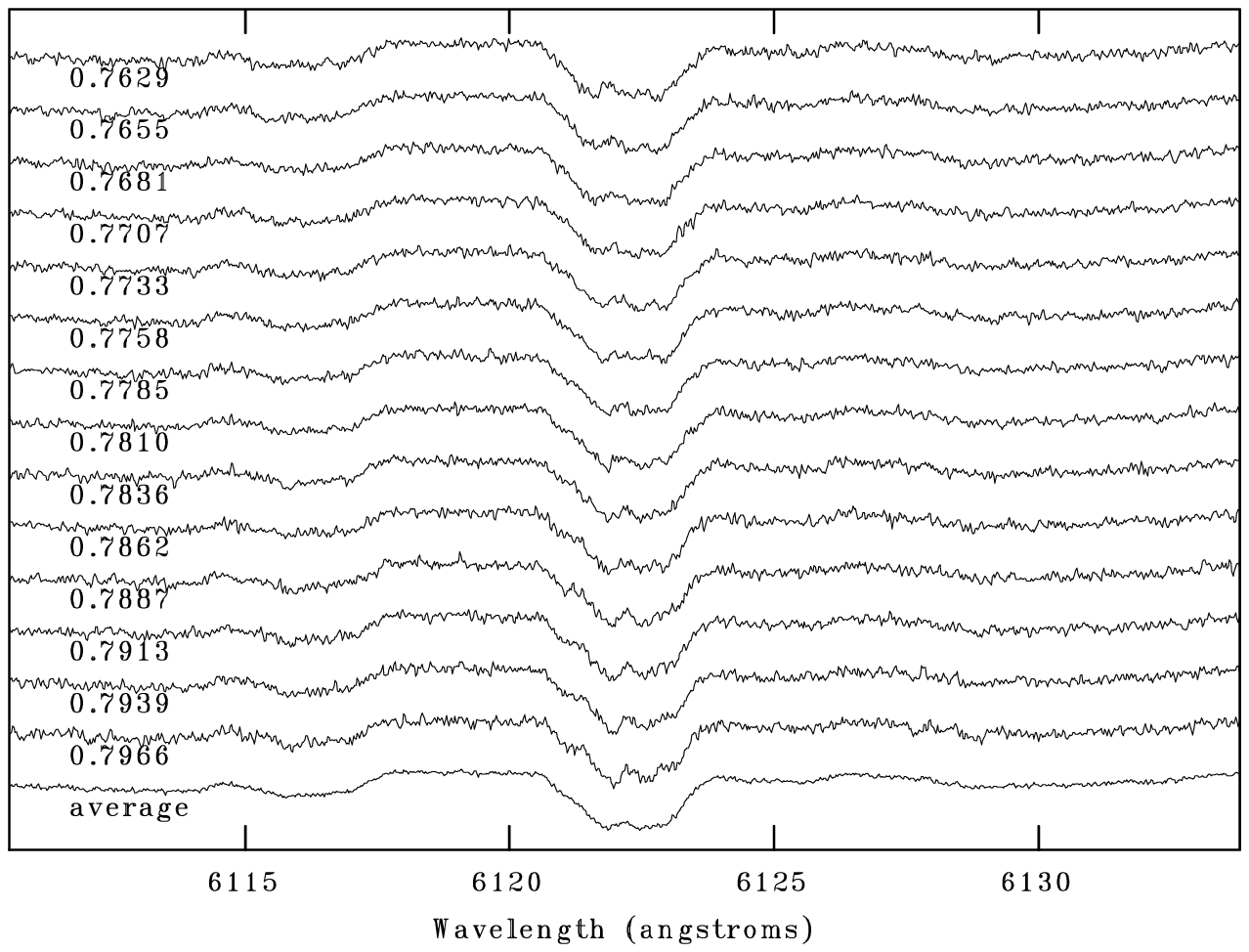}
    \end{minipage}
    \begin{minipage}[t]{8.5cm}
    \includegraphics*[width=9cm,height=7cm,bb=120 240 530 552]{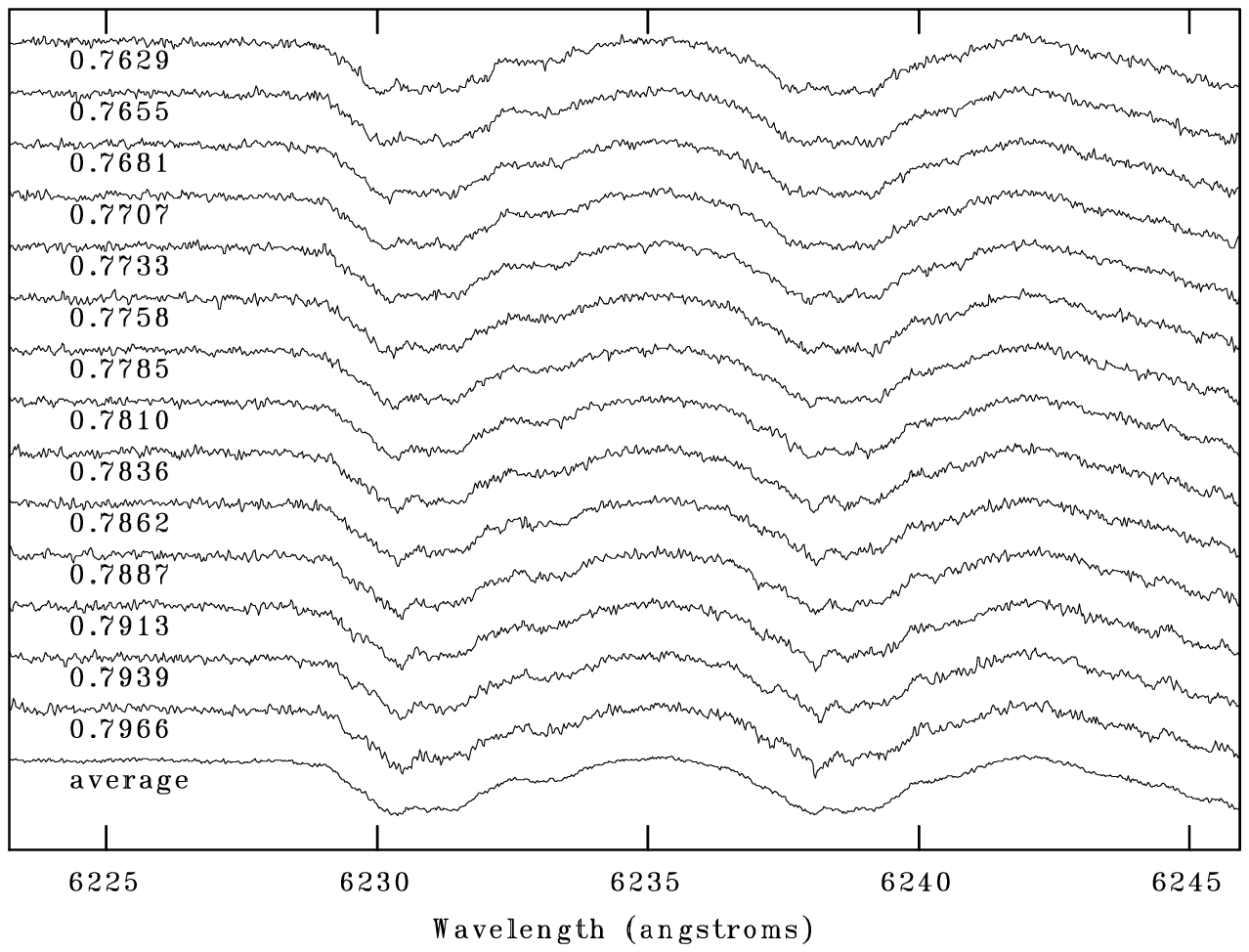}
    \end{minipage}
    \begin{minipage}[t]{8.5cm}
    \centering
    \hspace{-0.2cm} \includegraphics*[width=8.1 cm,height=1.0cm,bb=36 406 576 387]{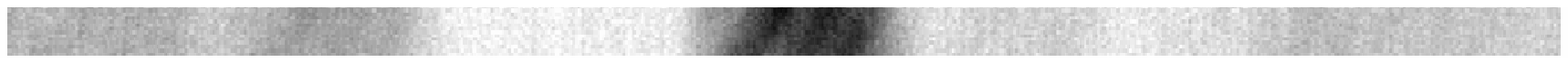}
    \end{minipage}
    \begin{minipage}[t]{8.5cm}
    \hspace{0.6cm} \includegraphics*[width=7.7 cm,height=1.0cm,bb=36 406 577 387]{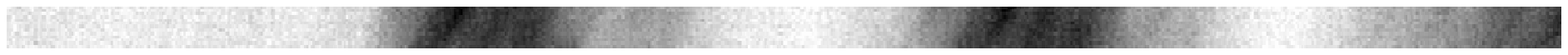}
    \end{minipage}

    \caption{Spectral profile variability of Mn\,\textsc{ii} 6122\,\AA{}, Fe\,\textsc{i}
6231\,\AA{}, and Fe\,\textsc{ii} 6238\,\AA{} lines in the UVES spectra of HD\,21190. Spectra are labelled with
Julian dates (HJD 2454047). The average spectra are plotted in the bottom of each panel.
The two-dimensional images present, for the same spectral regions, the 14 observed spectra stacked.
In these images the y-dimension corresponds to the observing time, increasing downwards.
Three moving peaks are clearly visible in the cores of all presented spectral lines.
}
  \end{center}
\end{figure*}

The obtained spectra show broad spectral lines ($v \sin i \sim 72$\,km\,s$^{-1}$), with small
variable peaks in the line profiles (Fig.\,1). In the bottom of Fig.\,1 
 we present all the
observed spectra stacked in two-dimensional images for two spectral regions. It can clearly be
seen that three peaks are moving smoothly towards the red with a speed of 21\,km\,s$^{-1}$ per hour.
However, no spectral features moving at higher frequencies were found in our spectra.
To measure the radial velocities of these peaks we filtered lower spectral frequencies and
used the cross-correlation method.

Similar splitting in the line profiles has been recently detected by Yushchenko et al. (2005)
in two other $\delta$\,Sct stars, $\delta$\,Sct itself and HD\,57749. Also, high signal-to-noise
spectroscopic time series observations by Poretti
for the $\delta$\,Sct star FG\,Vir, as reported by Mittermayer \& Weiss (2003), clearly show line
profile variations similar to those found in our spectra. These
are the typical blue-to-red moving bumps that are seen for non-radial pulsations
of non-axisymmetric m-modes in rapidly rotating stars where radial velocity
Doppler shifts the travelling waves of the individual pulsating surface segments,
the seminal example being the late O star $\zeta$\,Oph (Vogt \& Penrod 1983).

\section{The new rapidly oscillating Ap star HD\,218994}

HD\,218994 is a close visual binary system
with a separation of 1.2\,arcsec (Renson et al. 1991).
It is classified as A3\,Sr in the Michigan Spectral catalogue (Houk \& Cowley 1975).
Str\"omgren and H$\beta$ photometry were measured by Martinez (1993) who found
$V = 8.565$, $b - y = 0.154$, $m_1 = 0.196$, $c_1 = 0.826$ and $\beta = 2.807$.
Hubrig et al. (2000) estimated $T_{\rm  eff} = 7600$\,K from the $\beta$ index.
Martinez found $\delta m_1 = 0.008$ and $\delta c_1 = 0.032$ which are typical for a normal,
main sequence A star. He showed that correction for reddening increases the $m_1$ index by
about 0.02 and decreases the $c_1$ index by about 0.03. This star is unusual among the roAp
stars for its nearly-normal $m_1$ and $c_1$ indexes. Since most of the 36 known roAp stars
(see Kurtz et al. 2006 for a list) were found by photometric searches, this is an indication
that radial velocity studies will discover many more roAp stars that are difficult to detect
in photometric studies. HD\,218994 is located in the same region of the parameter space in which
rapid pulsations have been detected (Fig.\,2),  
but an earlier search for pulsations in the Cape Survey
yielded no detection (Martinez \& Kurtz 1994). We have several additional unpublished
photometric runs on this star that do not show roAp-type pulsations.

We obtained for this star 15 UVES spectra with a time resolution of $\sim$3 min.
From our spectra
we estimate $v \sin i = 5.2 \pm 0.6$\,km\,s$^{-1}$ using the first zero of the Fourier transform
of the spectral line profile of the magnetically insensitive lines.
Numerous lines of rare earth elements
(such as those of Nd\,\textsc{ii}, Nd\,\textsc{iii}, Pr\,\textsc{iii}, Eu\,\textsc{ii} and Tb\,\textsc{iii})
have been easily identified.
To search for pulsational line variability, we calculated the average spectrum of the observed
15 spectra and subtracted it from the original spectra.
We found for several rare earth lines a clear indication of variability.
This star is thus the 36$^{th}$ known roAp star.

\begin{figure}
\label{fig.hr}
  \begin{center}
    \includegraphics[height=8cm, width=8cm]{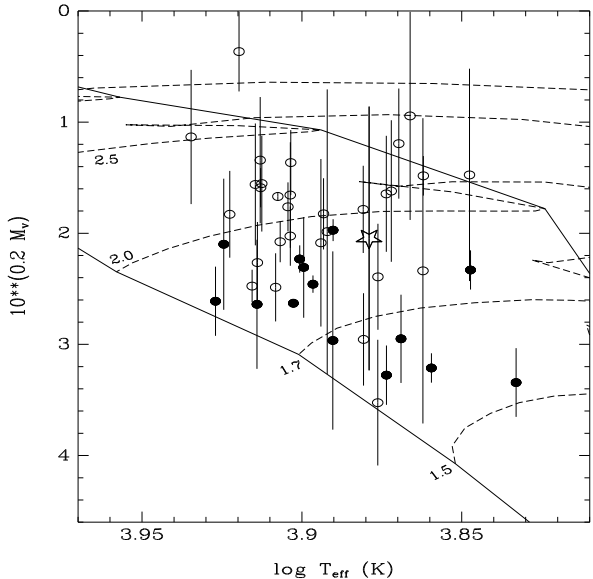}
    \caption{HR-diagram of the sample of roAp (full dots) and of the noAp stars (open dots)
studied by Hubrig et al. (2000).  HD 218994 is indicated by a star.  }
  \end{center}
\end{figure}

The lines of Fe\,\textsc{i}, Fe\,\textsc{ii}  and the Fe-peak elements show no radial velocity
variations with a 1$\sigma$ standard deviation per line in the range $40 - 70$\,m\,s$^{-1}$.
However, clear radial velocity variations are detected in the doubly ionised rare earth
elements lines of Nd\,\textsc{iii} and Pr\,\textsc{iii}, as is typical for the roAp stars
(see, e.g., Savanov, Maluneschenko \& Ryabchikova 1999; Kurtz, Elkin \& Mathys 2005).
We note that the mean radial velocity for different elements is different with a range
of a few km\,s$^{-1}$, indicating the presence of chemical inhomogeneities on the stellar surface.

We analysed 5 Nd\,\textsc{iii} lines, 8 Pr\,\textsc{iii} lines and a few lines of other
ions for radial velocity variations using a discrete Fourier transform (Kurtz 1985).
Individual lines of Nd\,\textsc{iii} and Pr\,\textsc{iii} show the same peak with a frequency
of 1.17\,mHz ($P = 14.2$\,min), confirming that the variation is in the star and is not instrumental.
Fig.\,3 
shows a radial velocity curve for 4 of the Nd\,\textsc{iii} lines averaged where
the 14.2-min pulsation is easily seen. Fig.\,4 
shows the amplitude spectra for
5 Nd\,\textsc{iii} lines and for 8 Pr\,\textsc{iii} lines
out to beyond the sampling frequency of 4.5 mHz.
Two peaks are seen, symmetrically placed on either side of the
Nyquist frequency. With the known range of pulsation periods
of roAp stars, $5.65-21.2$\,min, and our data sampling time of
3.7\,min, there is thus a potential ambiguity for the pulsation
period of HD\,218994. The Nyquist frequency for our data is
$\nu_{Ny} = 2.25$\,mHz.
As is expected for a purely sinusoidal variation and Gaussian noise, there is a peak
in Fig.\,4 at 3.3\,mHz ($\nu_S - \nu = 4.5 - 1.2 = 3.3$\,mHz; $P = 5.1$\,min).
See Kurtz (1983) for a graphical demonstration of this effect.

We consider it more likely that the 14.2-min period is the correct alias peak seen in Fig.\,4. 
But given that the shortest known period for an roAp star (HD\,134214) is only 5.65\,min,
we do not rule out the possibility that the 3.3\,mHz peak is the correct one. Higher time
resolution data will resolve this issue, as is usual with questions of Nyquist aliases.
Without another physical constraint on the choice of peak in Fig.\,4, 
either choice is  equally valid.
We consider higher frequency Nyquist aliases
to be physically implausible –-- the usual assumption for roAp stars.
Since each observational point represents the radial
velocity  averaged over the exposure time, the observed amplitude is reduced by a factor that
depends on the ratio between the exposure time and the period.
This effect reduces the observed amplitudes by a factor  0.52 for the 5.1 min period and
by a factor of 0.93 for the 14.1-min period. We thus expect higher
amplitudes to be measured with higher time resolution data.

Table~1 shows the least squares fit of the frequency 1.17\,mHz to the radial velocity
variations of some selected lines and for sets of Nd\,\textsc{iii} and Pr\,\textsc{iii} lines.
These results show that only Pr\,\textsc{iii} and Nd\,\textsc{iii} lines have significant
radial velocity variations, with amplitudes at least four times the standard deviation of the
measurements.
The data suggest, but do not prove, a higher amplitude for the
Pr\,\textsc{iii} lines than for the Nd\,\textsc{iii} lines.
There is no significant phase shift between them. In many roAp stars there appears to be
an outwardly running wave with Pr\,\textsc{iii} lines forming higher in the atmosphere
than Nd\,\textsc{iii} lines, hence  a phase shift between them.
It is often the case that the Pr\,\textsc{iii} lines show higher amplitude, as is expected
with the dropping atmospheric density. As our purpose here is to demonstrate unequivocally
that pulsation is present, and as the data are sparse, we defer further interpretation to
the acquisition of a more extensive data set.

\begin{figure}
\label{fig.rvNd}
  \begin{center}
\includegraphics*[width=7cm]{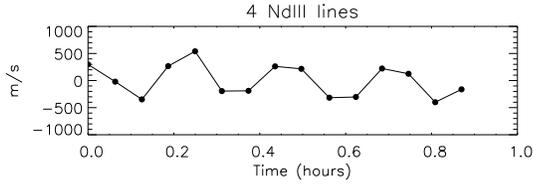}
    \caption{A radial velocity curve for the average of four Nd\,\textsc{iii} lines where
the 14.2-min pulsation can be clearly seen. The radial velocity error per observation is 220\,m\,s$^{-1}$. }
  \end{center}
\end{figure}

\begin{figure}
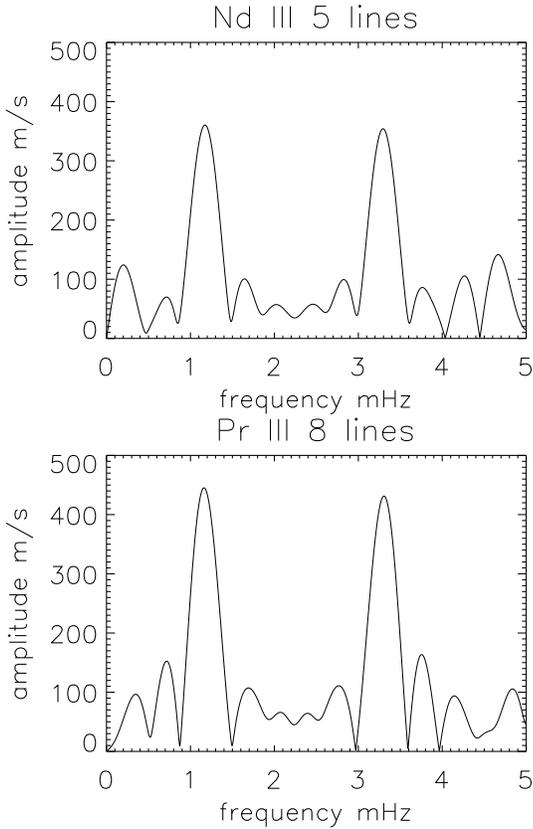

\label{fig.ampNdPr}
  \begin{center}
\includegraphics*[width=7cm]{ndiii5-ft.epsi}
\includegraphics*[width=7cm]{priii8-ft.epsi}
    \caption{Top panel: An amplitude spectrum for 5 Nd\,\textsc{iii} lines out to just
beyond the sampling frequency of 4.5 mHz. Two peaks are seen, as is expected, at 1.2 mHz
and 3.3 mHz, which is the sampling frequency minus 1.2 mHz.
Bottom panel: The same amplitude spectrum for 8 Pr\,\textsc{iii} lines
showing the same alias ambiguity. }
  \end{center}
\end{figure}

\begin{table*}
  \begin{center}
\label{table1}
    \caption[]{ A least squares fit of the frequency $\nu = 1.17$\,mHz to the radial velocity
variations of a selection of spectral lines of HD\,218994.  The third column gives the average
radial velocity  for each line; the range is indicative of horizontal abundance
variations. The zero point for phases is MJD\,54042.0.
The last columns give the standard deviation per observation of the 15 radial
velocity measurements to the fit, and the ratio between the amplitude and standard deviation
of the amplitude, i.e., the signal-to-noise in the detection of pulsation.}
    \begin{tabular}{ccccccc}
      \hline
      \multicolumn{1}{c}{$\lambda$} & \multicolumn{1}{c}{ion} &
      \multicolumn{1}{c}{RV} &  \multicolumn{1}{c}{amplitude}  &
      \multicolumn{1}{c}{phase} & \multicolumn{1}{c}{$\sigma$} & \multicolumn{1}{c}{amp/$\Delta$amp} \\
      \multicolumn{1}{c}{\AA} & \multicolumn{1}{c}{} &
      \multicolumn{1}{c}{km\,s$^{-1}$} & \multicolumn{1}{c}{m\,s$^{-1}$}  &
      \multicolumn{1}{c}{rad} & \multicolumn{1}{c}{m\,s$^{-1}$}  &
      \multicolumn{1}{c}{}\\
      \hline \\
5845.020 & Nd\,\textsc{iii} & $24.98\pm 0.09$ & $259 \pm 106$   & $-2.75 \pm 0.42$ & 292  &  2.4\\
5956.035 & Pr\,\textsc{iii} & $26.49\pm 0.08$ & $222 \pm \phantom{0}95$ & $\phantom{-}2.94 \pm 0.44$ & 265  & 2.3 \\
5987.683 & Nd\,\textsc{iii} & $24.88\pm 0.10$ & $457 \pm \phantom{0}66$ & $-2.67 \pm 0.15$ & 182  & 6.9 \\
6052.981 & Pr\,\textsc{iii} & $27.32\pm 0.10$ & $268 \pm 110$   & $-2.97 \pm 0.42$ & 305  & 2.4 \\
6089.985 & Pr\,\textsc{iii} & $26.64\pm 0.12$ & $360 \pm 127$ & $-2.87 \pm 0.36$ & 353  & 2.8\\
6141.713 & Ba\,\textsc{ii}  & $25.22\pm 0.04$ & $\phantom{0}71 \pm \phantom{0}51$ & $-1.17 \pm 0.71$ & 139  & 1.4 \\
6145.068 & Nd\,\textsc{iii} & $24.34\pm 0.04$ & $171 \pm \phantom{0}35$ & $-2.59 \pm 0.21$ & \phantom{0}95  & 4.9 \\
6147.741 & Fe\,\textsc{ii}  & $25.45\pm 0.03$ & $\phantom{0}91 \pm \phantom{0}39$ & $-1.13 \pm 0.42$ & 107  & 2.3 \\
6160.215 & Pr\,\textsc{iii} & $26.51\pm 0.22$ & $931 \pm 178$ & $-2.48 \pm 0.19$ & 490  & 5.2 \\
6195.611 & Pr\,\textsc{iii} & $25.52\pm 0.14$ & $510 \pm 116$ & $-2.07 \pm 0.23$ & 304  & 4.4 \\
6327.265 & Nd\,\textsc{iii} & $25.15\pm 0.13$ & $600 \pm \phantom{0}93$ & $-2.75 \pm 0.16$ & 258  & 6.5 \\
6550.231 & Nd\,\textsc{iii} & $24.84\pm 0.07$ & $299 \pm \phantom{0}58$ & $-2.93 \pm 0.20$ & 162  & 5.2 \\
6562.797 & H$\alpha$        & $25.11\pm 0.12$ & $121 \pm \phantom{0}57$ & $-1.30 \pm 0.46$ & 155  & 2.1 \\
6645.064 & Eu\,\textsc{ii}  & $24.96\pm 0.09$ & $205 \pm 107$ & $\phantom{-}0.99 \pm 0.52$ & 293  & 1.9 \\
6910.146 & Pr\,\textsc{iii} & $24.92\pm 0.14$ & $352 \pm 172$ & $\phantom{-}2.90 \pm 0.50$ & 477  & 2.0 \\
7030.388 & Pr\,\textsc{iii} & $25.22\pm 0.13$ & $586 \pm \phantom{0}94$ & $-2.76 \pm 0.17$ & 262  & 6.2 \\
7076.618 & Pr\,\textsc{iii} & $25.12\pm 0.14$ & $511 \pm 138$ & $-2.78 \pm 0.28$ & 383  & 3.7 \\
5 lines  &  Nd\,\textsc{iii}&  & $355 \pm \phantom{0}38$ & $-2.74 \pm 0.11$ & 238  & 9.3 \\
8 lines  & Pr\,\textsc{iii} &  & $437 \pm \phantom{0}53$ & $-2.73 \pm 0.12$ & 412  & 8.2 \\
\hline
\end{tabular}
\end{center}
\end{table*}

\section{Conclusions}

We report the discovery of a new roAp star, HD 218994, with a pulsation period of 14.2 min, or 5.1\,min.
This is one of the most evolved of all roAp stars, hence is an important test of theoretical
models of pulsation driving and pulsation periods in roAp stars.
No rapid pulsations have been
found in the spectra of the cool Ap -- $\delta$ Scuti star HD\,21190, hence it is not an roAp star.
However, we detect moving peaks
in the cores of spectral lines  -- with a clarity never seen before in a $\delta$\,Sct star --
which indicate the presence of non-axisymmetric
non-radial pulsations in this star. Its combination of cool Ap spectral type and high degree $\delta$\,Sct
pulsation makes it an important target for more in depth study.


{}


\begin{thebibliography}{}

\bibitem[2000]{pipe} Ballester, P., Grosbol, P., Banse, K., et al., 2000, SPIE, 4010, 246

\bibitem[2001]{bal01} Balmforth, N. J., Cunha, M. S., Dolez, N.; Gough, D. O., Vauclair, S., 2001, MNRAS, 323, 362

\bibitem[2002]{cu02} Cunha, M. S.,  2002, MNRAS, 333, 47

\bibitem[2005]{el05} Elkin, V. G., Riley, J. D., Cunha, M. S., Kurtz, D. W., Mathys, G., 2005, MNRAS, 358, 665

\bibitem[1975]{michi} Houk, N., Cowley, A. P., 1975, Michigan Catalogue of two-dimensional spectral types for the HD star, Ann Arbor: University of Michigan

\bibitem[2000]{hu00} Hubrig, S., Kharchenko, N., Mathys, G., North, P., 2000 A\&A,  355, 1031

\bibitem[2001]{ko01} Koen, C., Kurtz, D. W., Gray, R. O., Kilkenny, D., Handler, G., Van Wyk, F., 2001, MNRAS, 326, 387

\bibitem[1983]{ku83} Kurtz D.~W., 1983, IBVS, 2285, 1

\bibitem[1985]{ku85} Kurtz, D. W., 1985, MNRAS, 213, 773

\bibitem[2006]{ku06} Kurtz D. W., Elkin, V. G., Cunha, M. S., Mathys, G., Hubrig, S., Wolff, B., Savanov, I., 2006, MNRAS, 372, 286

\bibitem[2005]{kem05} Kurtz, D. W., Elkin, V. G., Mathys, G., 2005,  2005, EAS, 17, 91 

\bibitem[1993]{ma93} Martinez, P., 1993, PhD Thesis, University of Cape Town

\bibitem[1994]{ma94} Martinez, P., Kurtz, D. W., 1994, MNRAS, 271, 129

\bibitem[2003]{mitt} Mittermayer, P., Weiss, W., 2003, A\&A, 407, 1097

\bibitem[2002]{asas} Pojmanski, G., 2002, Acta Astronomica, 52, 397

\bibitem[1991]{re91} Renson, P., Gerbaldi, M., Catalano, F. A., 1991, A\&As, 89, 429

\bibitem[2005]{sai} Saio, H., 2005, MNRAS, 360, 1022

\bibitem[1999]{sa99} Savanov, I. S., Malanushenko, V. P., Ryabchikova, T. A., 1999, Ast.L., 25, 802

\bibitem[1983]{vp83} Vogt, S. S., Penrod, G. D., 1983, ApJ, 275, 661

\bibitem[2005]{yu05} Yushchenko, A., Gopka, V., Kim, Chulhee, Musaev, F., Kang, Y. W., Kovtyukh, V., Soubiran, C., 2005, MNRAS, 359, 865

\end{thebibliography}
\end{document}